\documentclass[10pt]{iopart}
\usepackage{amssymb}
\usepackage{soul}
\usepackage{graphicx}
\usepackage{framed}

\usepackage{geometry}
\usepackage{marginnote}
\usepackage[switch,columnwise]{lineno}

\usepackage[export]{adjustbox}
\usepackage[justification=centering]{caption}
\usepackage[font={footnotesize,it}]{caption}
\usepackage{cite}
\usepackage{color}
\usepackage{xcolor}
\usepackage{bm}
\usepackage{hyperref}
\usepackage{colortbl}
\newcommand{\ket}[1]{|#1\rangle}
\newcommand{\bra}[1]{\langle #1|}

\usepackage{xcolor,colortbl}
\definecolor{Gray1}{gray}{0.8}
\definecolor{Gray2}{gray}{0.9}
\begin{document}

\title{Exact dynamics of concurrence-based entanglement in a system of four spin-1/2 particles on a triangular ladder structure}

\author{Sajedeh Shahsavari${}^{1}$, Mostafa Motamedifar${}^{2}$, and 
 Hassan Safari${}^{1}$}
\address{${}^1$ Department of Photonics, Faculty of Modern Science and Technologies, Graduate University of Advanced Technology, Kerman, Iran}
\address{${}^2$ Faculty of Physics, Shahid Bahonar University of Kerman, Kerman, Iran}

\begin{abstract}
Motivated by the ability of triangular spin ladders to implement quantum information processing, we propose a type of such systems whose Hamiltonian includes the $XX$ Heisenberg interaction on the rungs and Dzyaloshinskii–Moriya (DM) coupling over the legs.
In this work, we discuss how tuning the magnetic interactions between elements of a nanomagnetic cell of a triangular ladder which contains four qubits influences on the dynamical behavior of entanglement shared between any pairs of the system. In this work, we make use of concurrence for monitoring entanglement.
It is realized that the generation of quantum $W$ states is an important feature of the present model when the system evolves unitarily with time. 
In general, coincidence with the emergence of $W$ states, the concurrences of all pairs are equal to $2/N$, where $N$ is the number of system's qubits. 
We also obtain the precise relationship between the incidence of such states and the value of DM interaction as well as the time of entanglement transfer.‌‌‌‌‌
Finally, by studying the two-point quantum correlations and expectation values of different spin variables, we find that $xx$ and $yy$ correlations bring the entanglement to a maximum value for $W$ states, whereas for these states, $zz$ correlation between any pairs completely quenches. Our results reveal that although $\hat{S}^{z}_{tot}$ does not commute with the system's Hamiltonian, its expectation value remains constant during time evolution which is a generic property of quantum $W$ states. 

\vspace{2ex}\noindent
\end{abstract}


%
\ioptwocol

\section{Introduction}
Quantum state transfer is one of the crucial demands to perform quantum information processing. Depending on the accessible technology, different methods exist to accomplish the transfer of quantum states and consequently the propagation of quantum entanglement.  As an illustration, in optical systems, this task is attributed to photons~\cite{guthohrlein2001single, jons2017bright}. Moreover, in the systems of trapped atoms, phonons play the role of information carriers~\cite{ruskov2012coherent}. 

The desire to achieve greater processing power in quantum computation makes quantum bits (qubits) get closer enough to each other for  more feasible transferring quantum states. Then, such a necessity induces an additional size limitation on quantum communication devices. Short range quantum communication called for promising devices to realize transferring a quantum state from one location to another one in short distances. Meanwhile, focusing on time evolution of quantum spin chains has switched into high gear for such a goal, after Boes's idea~\cite{bose2003quantum, bose2007quantum}. 

Bose suggested that quantum state propagation can be fulfilled by the dynamic evolution of an array of permanently coupled spins. In his suggested scenario various aspects come into play: 
decoding a quantum state at the $s$-th spin of a chain by Alice, travelling the state through the chain and retrieving some fidelity of the mentioned state from $r$-th spin after a while by Bob. 
It was found that the amount of fidelity is equal to $1$ for a four-spin ring which denotes such a ring can provide a perfect quantum communication~\cite{bose2003quantum}. 
In addition, magnetic spin chains meet many other requirements in several inspiring technologies for the processing of quantum information~\cite{apolaro2015, simon2011quantum, blanc2018quantum, bazhanov2018engineering}.

It is noticed in Ref.~\cite{wang2001 } that during the evolution of a type of spin chain over time, the unique quantum states of $W$ type can be dynamically produced. 
Such states can be classified among the well-known multipartite quantum states that play a significant role in the quantum information tasks~\cite{dur2000}. Tracing out one qubit of the system results in another $W$ sate. In other phrase, such states are not sensitive to the loss of particles. The general form of quantum $W$ states can be represented by 
\begin{eqnarray}
|W\rangle^N=\frac{1}{\sqrt{N}}\sum_{j}^{N}e^{i\theta_j}\ket{\bf j},
\label{eq:wstates}
\end{eqnarray}
where $|{\bf j}\rangle=|00...010....0\rangle$ presents a state with a flipped spin at the $j$-th site to the $|1\rangle$ state. While $|0\rangle$ indicates the spin down state (i.e., spin oriented along $-z$ orientation), $|1\rangle$ denotes the reversed direction. Also, $\theta_j$ is an arbitrary phase factor.

Factually, such states are an essential type of physical resources for lots of notable applications such as  quantum key distribution~\cite{cao2006quantum, li2009improvement}, quantum telecloning~\cite{murao1999quantum}, quantum teleportation~\cite{shi2002teleportation, dai2003probabilistic, xiao2008quantum}, just to name a few. Above all, the generation of $W$ states is the primary condition of such captivating applications. From dynamical point of view, the quantum $W$ states can be generated via evolution of an initial quantum state either in a spin chain~\cite{wang2001} or in a spin star network~\cite{motamedi20171}. It may be interesting that, opposing to spin chains through which $W$ states can be dynamically created by four qubits at the most~\cite{wang2001}, star networks have shown that such states can be generated by
the same number of qubits at the least~\cite{motamedi20171}.

In what follows, we show how such states can be generated during the evolution of our considered system i.e.,  a four-qubit triangular nanomagnetic ladder. 
If two spin chains are arranged parallel (or quasi-parallel) to each other in a two-dimensional plane, it is known to result in systems called two-leg ladders and can be used to implement quantum channels.
A deep insight into the physics of such quantum channels is of particular importance for the technologies associated with the entanglement and quantum state transfer on them. These technologies covers a range from the miniaturized structures such as quantum cellular automata~\cite{elze2013quantum} to the larger-sized components as quantum computer's data buses~\cite{jia2015integrated}.

We assumed that in such a system, spins are coupled by the Heisenberg interaction of strength $J$ along the ladder's rungs as well as being interacted by spin-orbit coupling over the legs of ladder.
This proposed model is of profound importance for the generation of quantum $W$ states because, the two types of magnetic interactions coexist in such a spin arrangement beyond the spin chains or star structures that are customarily well known in the generation of $W$ states.

There are several suggestions for engineering spin-orbit coupling for atoms in optical lattices, one of the most significant of which is based upon the superexchange mechanism of Dzyaloshinskii–Moriya (DM) effect~\cite{dzya, moriya}. Such an interaction leads to emergence of exotic supperfluidity~\cite{tobias} and gives rise to a broad class of spin patterns in quantum phases~\cite{messio2017, strag2017chiral, Lee2018spinhall, jafari2018} and entails a dominant form of anisotropy in spin liquids~\cite{balentsspinliquid2019} just to name a few. The anisotropy parameter is important because it can broaden the line-with of electron spin resonance in kagome and honeycomb lattices as well as triangular ones with spin-orbit coupling that are the most promising candidates for spin liquids to emerge~\cite{jpcm2019}.

In this work, we demonstrate how tuning DM interaction in a four-qubit spin ladder influences on the dynamical behavior of entanglement shared between any pairs of the system. This objective is fulfilled by means of concurrence~\cite{hill1997entanglement, wootters1998entanglement, coffman2000distributed} to monitor the dynamical procedure. 
Recently, the focus has been on developing the role of concurrence in many body systems as well as in the processing of quantum information. For instance, in solid state physics, characterizing some phases of several materials is devoted to this quantity~\cite{azimi2016pulse}. In addition to its interest in analytical approaches, several numerically effective methods for calculating this quantity have been developed as well~\cite{vimal2018quantum}. As it known, one of the situations in which the concurrences shared on any pairs of a system are all equal is related to the quantum $W$ states~\cite{zhu2011, liu2012}.

For dealing with the above situations in details, our paper is organized as follows: In Sec.~\ref{sec:model} we introduce the considered model and describe how the initial state of the system evolves during the time intervals. The beginning of Sec.~\ref{sec:concu} is devoted to recall concurrence as the considered measure tool for qubit-qubit entanglement.  The analytical results presented in this section have provided the exhaustive descriptions of the dynamical behavior of entanglement between any pairs. Section~\ref{sec:twopointcorre} addresses the two point correlation functions and their relations with entanglement. In the concluding section we give a summary of our results.

\section{Description of the Model}
\label{sec:model}
Let us consider a ladder in which spin 1/2 particles interact via XX Heisenberg Hamiltonian on the rungs.
In addition, DM interaction links spins over the legs.
The Hamiltonian that describes such a system with assuming a periodic boundary condition on the rungs, reads:
\begin{figure}
	\centering
	\includegraphics[width=0.4\linewidth]{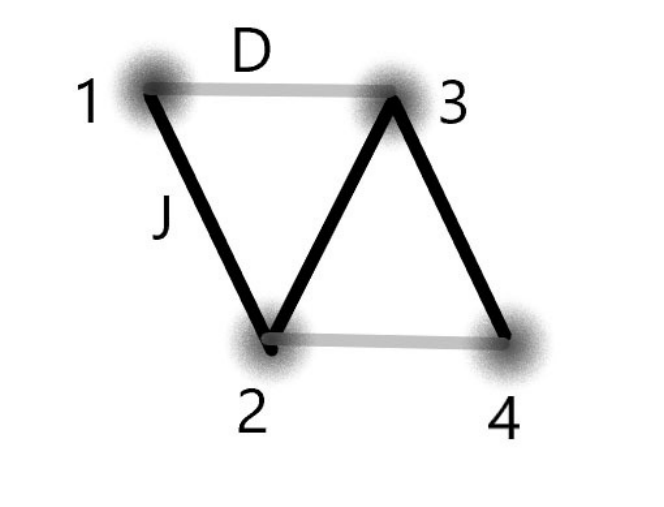}
	\caption{Nomination of the sites on the ladder structure. Two horizontal links (between pairs ($1,3$) and ($2,4$)) in the shape are called legs here and other links are rungs.}
	\label{fig:Figladder}
\end{figure}

\begin{eqnarray}
\hat{H}=J\sum_{<i,j>}(\hat{S}^{x}_{i}\hat{S}^{x}_{j}+\hat{S}^{y}_{i}\hat{S}^{y}_{j})+\vec{D}
\boldsymbol{\cdot}\sum_{\ll i',j'\gg}\hat{\vec{S}}_{i'}\times\hat{\vec{S}}_{j'},
\label{Ham}
\end{eqnarray}
where $<i,j>$ $\big(\ll i',j'\gg\big)$ denotes sites on the  rungs (legs). In this equation, the
first term in the right hand side stands for $XX$ Heisenberg interaction with the parameter $J$ showing its strength on the rungs. 
The second term takes into account the DM interaction (as depicted in Fig.~\ref{fig:Figladder}) over the legs. Also, $\vec{D}=D\hat{z}$ plays the control parameter's role in the system dynamics. 

In order to study the dynamics of the system, we make use of the time evolution operator approach for which the initial quantum state of the system is required.
As long as the system undergoes a unitary dynamics generated by the system's Hamiltonian due to the time evolution operator $\hat{U}(t) = \exp(-i \hat{H} t/\hbar)$ we wish to examine its application on the initial state which can be written as
\begin{eqnarray}
&&\hat{U}(t)\ket{\psi(0)}=\exp\bigg\{\frac{-iJt}{\hbar}\bigg[\sum_{<i,j>}(\hat{S}^{x}_{i}\hat{S}^{x}_{j}+\hat{S}^{y}_{i}\hat{S}^{y}_{j})\nonumber\\
&&\qquad+\frac{\vec{D}}{J}
	\boldsymbol{\cdot}\sum_{\ll i',j'\gg}^{2}\hat{\vec{S}}_{i'}\times\hat{\vec{S}}_{j'}\bigg]\bigg\}\ket{\psi(0)}.
\label{Hamdim}
\end{eqnarray}
In what follows, $t$ and $D$ respectively are normalized to $\hbar/J$ and $J$. Without loss of generality, we take $\hbar\equiv1$ and $J=1$ here. Therefore, $t$ and $D$ become dimensionless parameters in the following computations.

Let's consider the case in which the entanglement is initially encoded in the entangled Bell state $\frac{1}{\sqrt{2}}(\ket{10} _{12} + \ket{01}_ {12})$ over the first two qubits (1,2),
while the last two qubits (3,4) are in the disentangled state $\ket{00}_{34}$. 
Hence, the initial state of the whole system can be shown by a tensor product state, $\ket{\psi(0)}=\frac{1}{\sqrt{2}}(\ket{10} _{12} + \ket{01}_ {12})\ket{00}_{34}$.

In order to proceed Eq.~(\ref{Hamdim}), $\ket{\psi(0)}$ is spanned over energy eigenstates. 
Factually, we can write $|\psi(0)\rangle$ in term of energy eigenvectors ($|E_i\rangle$) as
\begin{eqnarray}
|\psi(0)\rangle=\sum_{i=1}^{16}c_i|E_i\rangle.
\end{eqnarray}
To reach $|\psi(t)\rangle$ we apply the time evolution operator as
\begin{eqnarray}
|\psi(t)\rangle=e^{\frac{-i Ht}{\hbar}}|\psi(0)\rangle=\sum_{i=1}^{16}c_ie^{\frac{-i E_i t}{\hbar}}|E_i\rangle.
\label{eq:unitapli}
\end{eqnarray}
Moreover, we know that it is possible to extend energy states onto standard bases as $|E_i\rangle=\sum_{j=1}^{16} a_j |\phi_j\rangle$ which can be inserted in Eq.~\ref{eq:unitapli}. Here $|\phi_j\rangle$ are included in the set of standard bases $\{|1111\rangle, |1110\rangle,...|0000\rangle\}$. Then we have 
\begin{eqnarray}
|\psi(t)\rangle=\sum_{i,j=1}^{16}c_i a_j e^{\frac{-i E_i t}{\hbar}}|\phi_j\rangle.
\label{eq:unitapli2}
\end{eqnarray}
After straightforward calculation, it is found that all coefficients emerge in Eq.~\ref{eq:unitapli2} are zero except those which are related to \textit{one-particle} state~\cite{wang2001}. In such states, one particle is up and the rest are down in the $z$ direction of spin space.  Consequently, the time evolution of the quantum state reads:
\begin{eqnarray}
&&\ket{\psi(t)}=\frac{\eta{(t, D)}}{2\sqrt{2}}(\ket{1000}+\ket{0100})\nonumber\\
&&\qquad+\frac{\xi{(t, D)}}{2\sqrt{2}}(\ket{0010}+\ket{0001}),
\label{kett}
\end{eqnarray}    
where
\begin{eqnarray}
&&\eta(t,D)=\cos\frac{\mu t}{2}+
\cos\frac{\nu t}{2}\nonumber\\
&& \qquad-\frac{i}
{\omega}\left[\frac{\omega+\omega^2}{\mu}\sin\frac{\mu t}{2}
+\frac{\omega-\omega^2}{\nu} \sin\frac{\nu t}{2}\right],
\nonumber\\
&&\xi(t,D)=\frac{-(i+D)}{\omega D^{2}}\left\{i D^2\left[
\cos\frac{\mu t}{2}-\cos\frac{\nu t}{2}\right]\right. 
\nonumber\\
&&\qquad\left.+(\omega+1)\nu \sin\frac{\mu t}{2} 
+(\omega-1)\mu\sin\frac{\nu t}{2}\right\},
\label{eq3x}
\end{eqnarray}
in which $\omega = \sqrt{1+D^2}$ , and $\mu$ and $\nu$ are 
positive parameters depending on $D$ as:
\begin{eqnarray}
&&\mu =\sqrt{2+D^{2}+2\sqrt{1+D^{2}}},\nonumber\\
&&\nu =\sqrt{2+D^{2}-2\sqrt{1+D^{2}}}.
\end{eqnarray}
As we can see in Eq.~(\ref{kett}), the quantum state for an arbitrary time instant is a linear combination of the states in which the qubits pair 
(1,2) is in an entangled Bell-state while the pair (3,4) is in the ground state, and the other way around.
In what follows, we make use of Eq.~\ref{kett} to calculate the value of entanglement between different parts of the considered system at different time instants.

\section{Entanglement dynamics}
\label{sec:concu}
As suggested by Wootters et al.~\cite{hill1997entanglement, wootters1998entanglement, coffman2000distributed}, the concurrence is an appropriate quantifier for the pairwise entanglement between qubits $p$ and $q$ in a quantum system. The starting point is tracing of the density matrix $\rho$ over two qubits $p$ and $q$, which results in the reduced density matrix $\rho(p,q)=\Tr_{p,q}(\rho)$. Then, the concurrence between $p$-th qubit and $q$-th one is given by $C_{p,q}=max\{2\lambda_1-\sum_{i=1}\lambda_i,0\}$ where $\lambda_i$s are square roots of eigenvalues of the matrix $R=\rho(p,q)\big(\sigma_y\otimes\sigma_y\big)\rho^{*}(p,q)\big(\sigma_y\otimes\sigma_y\big)$. Moreover, $\sigma_y$ denotes the Pauli-y-matrix and $\rho^{*}(p,q)$ is complex conjugate of $\rho(p,q)$. 
In definition,
$0 < C_{p,q} < 1$ indicates that $p$-th qubit is partially entangled with $q$-th one. While, $C_{p,q} = 1$ corresponds to a maximally entangled state, $C_{p,q} = 0$ stands for a completely disentangled state between two considered qubits. 
While, for a through discussion and comparison between concurrence and other quantum correlations we refer for instance to Ref.~\cite{mot2017} and bibliography quoted therein.

The concurrence, for a quantum state that includes one-particle states as relation.\ref{kett}, can be simplified by $C_{p,q}(t) = 2|b_p(t)b_q(t)|$ where, $b_i(t)$s are the coefficients in that relation~\cite{wang2001, amico2004}.
For the considered system, our calculation shows that the concurrences are given by $C_{1,2}=\cos^{2}(\frac{\mu+\nu}{4}t)$ and $C_{3,4}= \sin^{2}(\frac{\mu+\nu}{4}t)$ for the first and the last rung respectively. Moreover, for every leg we have $C_{1,3}=\frac{1}{2}|\sin(\frac{\mu+\nu}{2}t)|$. It might be worthy to mention that $C_{2,3}$ and $C_{1,4}$ behave as $C_{1,3}$.
 
\begin{figure*}[t]
	\centering
	\begin{tabular}[b]{c}
		\includegraphics[width=.4\linewidth]{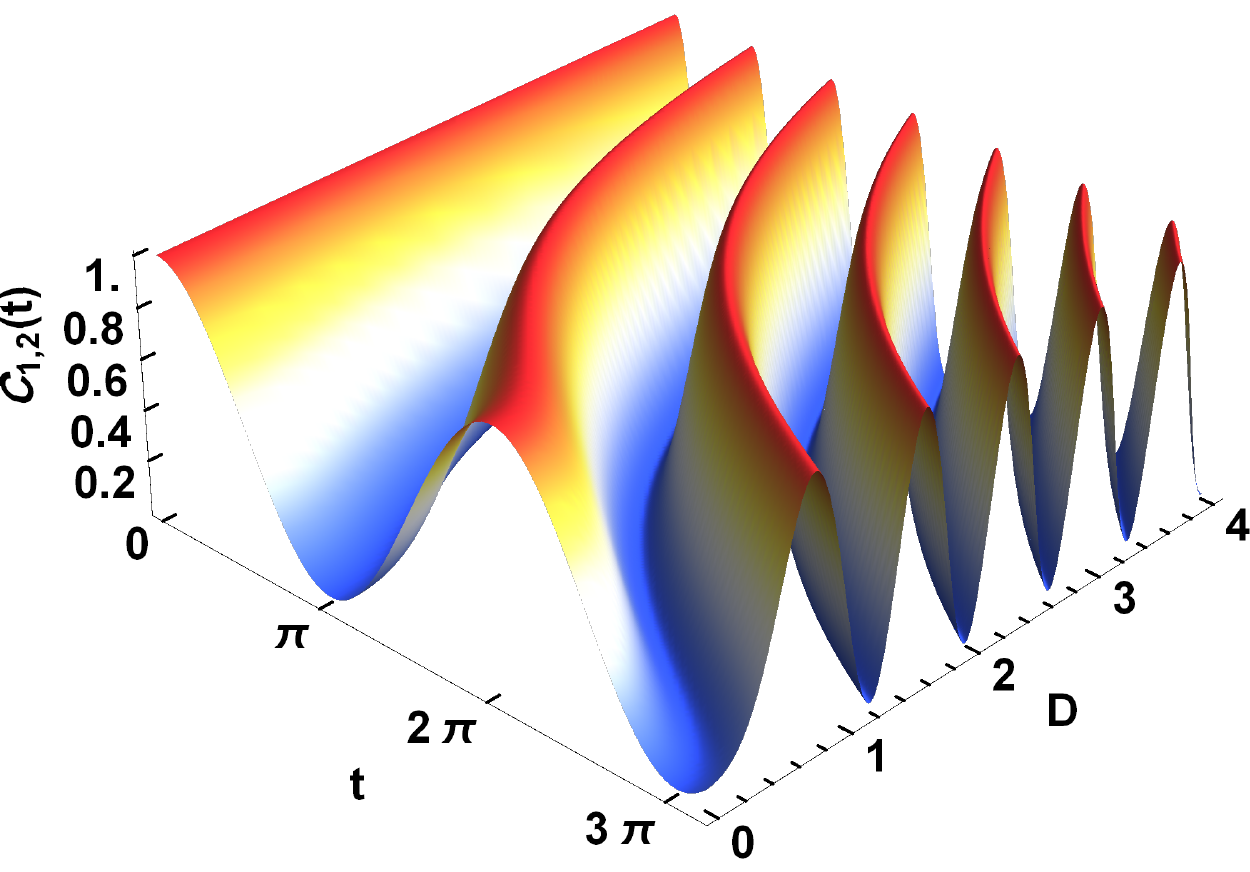}\\
		\small (a)
	\end{tabular}
	\begin{tabular}[b]{c}
		\includegraphics[width=.4\linewidth]{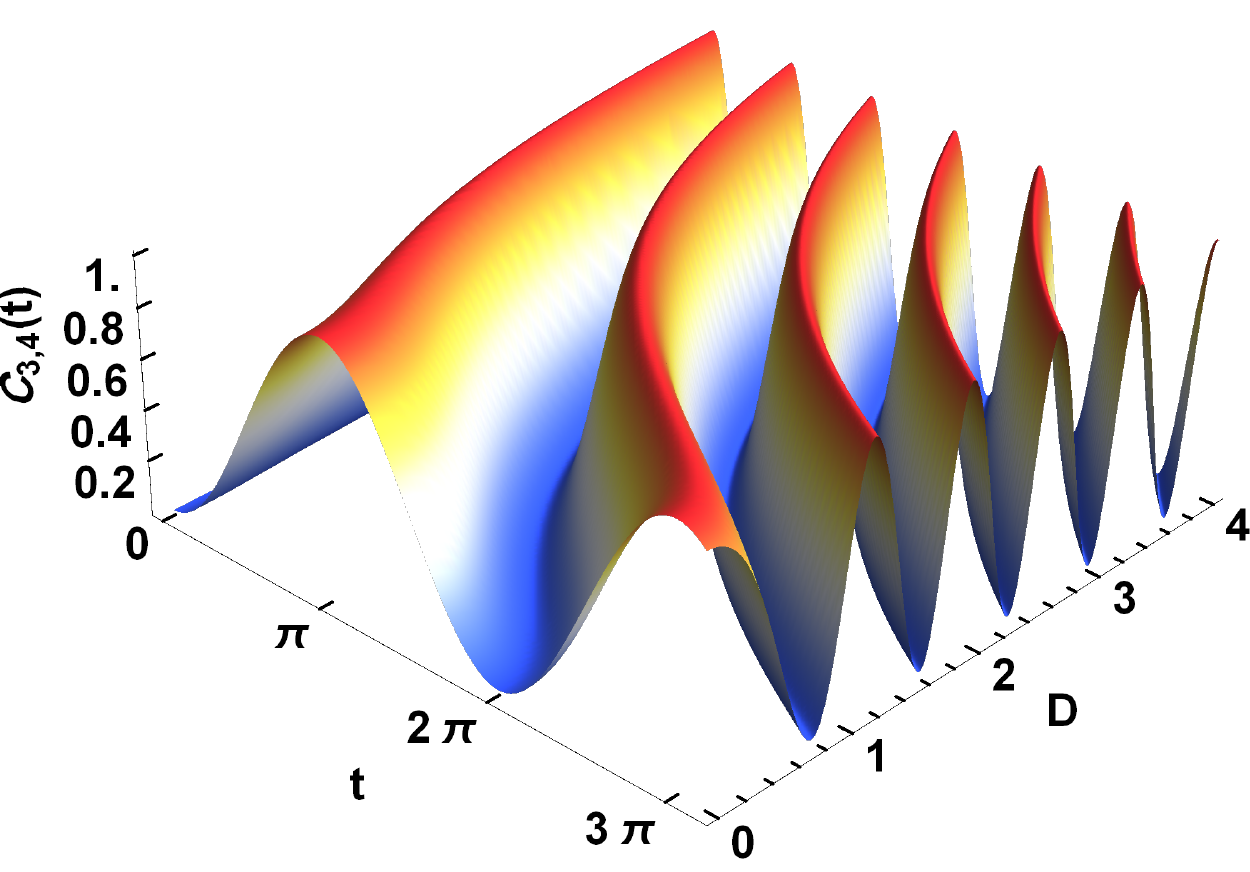}\\
		\small (b)
	\end{tabular} 
	\begin{tabular}[b]{c}
		\includegraphics[width=.4\linewidth]{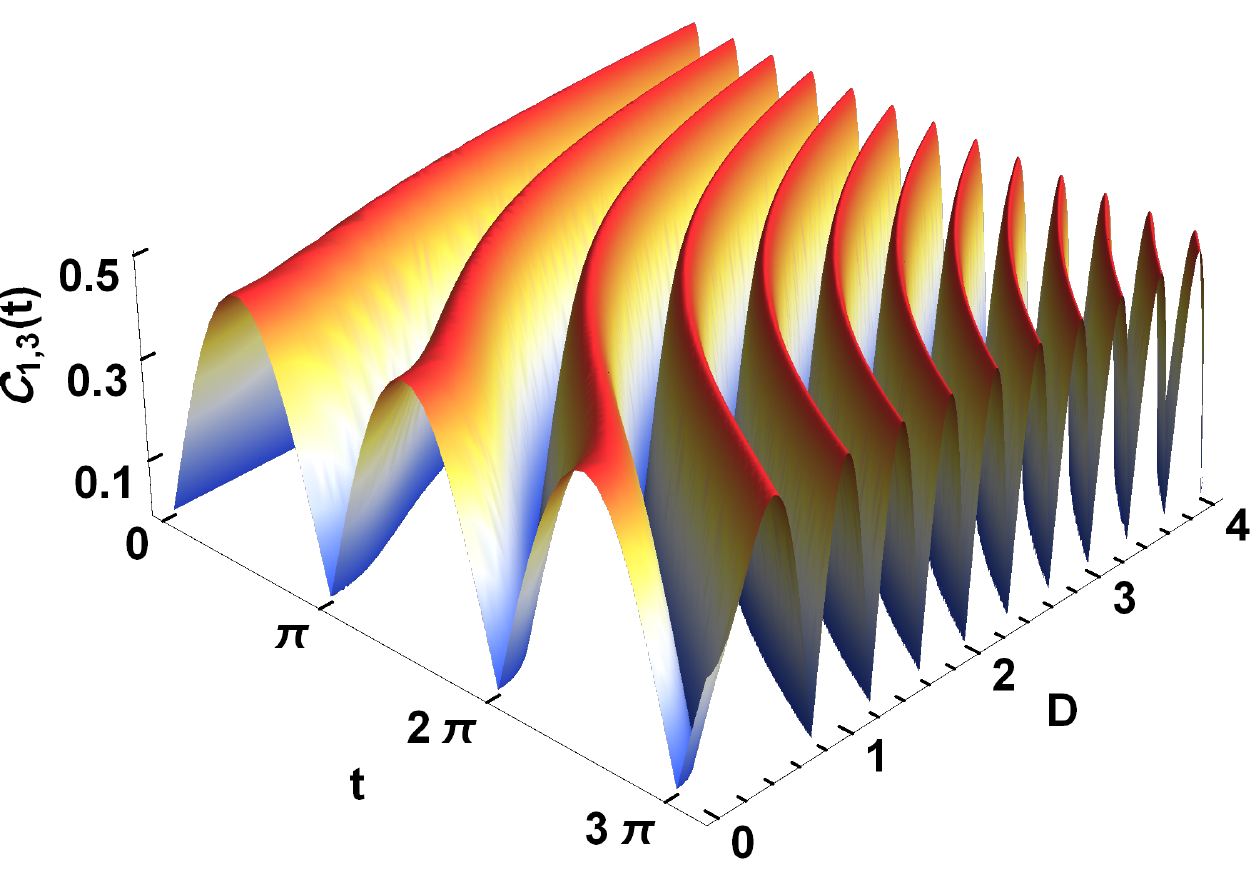}\\
		\small (c)
	\end{tabular} 
	\caption{ Three dimensional plots of entanglement oscillation as a function of 
		time and $D$ for (a)~the first pair, (b)~the last pair, (c)~between pair $(1,3)$ which is the representative of other pairs.}
	\label{fig:Fig13D}
\end{figure*}

Figures~\ref{fig:Fig13D}~(a) and~(b)
represent the dynamical behavior of the concurrence, respectively, between the first and last leg in terms of the DM interaction parameter $D$. Sharafullin et~al. demonstrated in details how DM Hamiltonian relates to the angle between spins and the value of nonmagnetic ions (for example oxygen) displacement in a materials~\cite{shaff2019}. 
However, other possible schemes might come into play for determining DM interaction. For instance, temperature and voltage bias applied to the system and spin polarization can effect the DM interaction~\cite{fransson2014}. In Ref.~\cite{kato2019}, it is shown that the interfacial DM interaction is not a given material parameter. Instead, it can be controlled and managed by a spin current injected externally to the system. In any way, without concerning about the experimental method to control DM interaction, we consider them theoretically in a range of values.
\\
\begin{figure*}[h!]
	\begin{tabular}[b]{c}
		\includegraphics[width=0.45\linewidth]{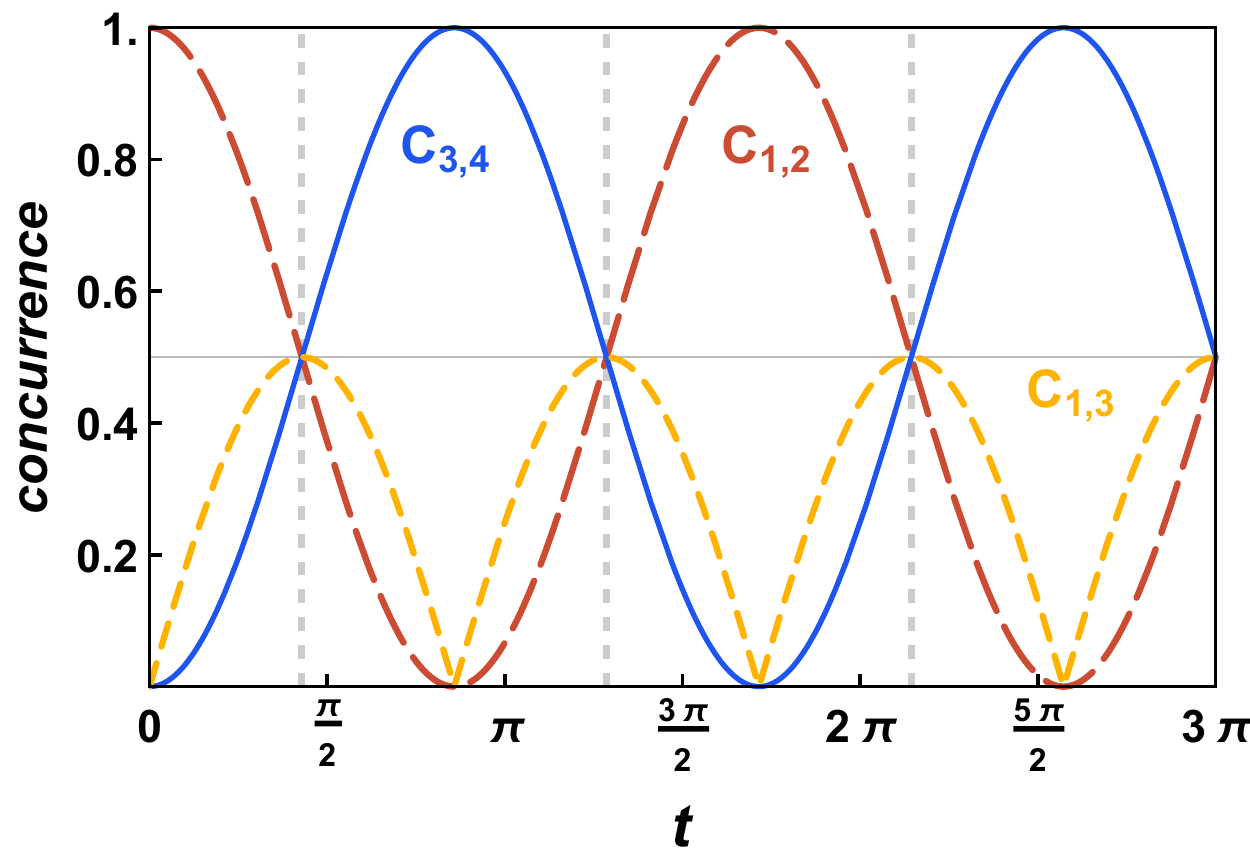}\\
		\label{fig:fig2D}
		\small (a)
	\end{tabular}
	\begin{tabular}[b]{c}
		\includegraphics[width=0.45\linewidth]{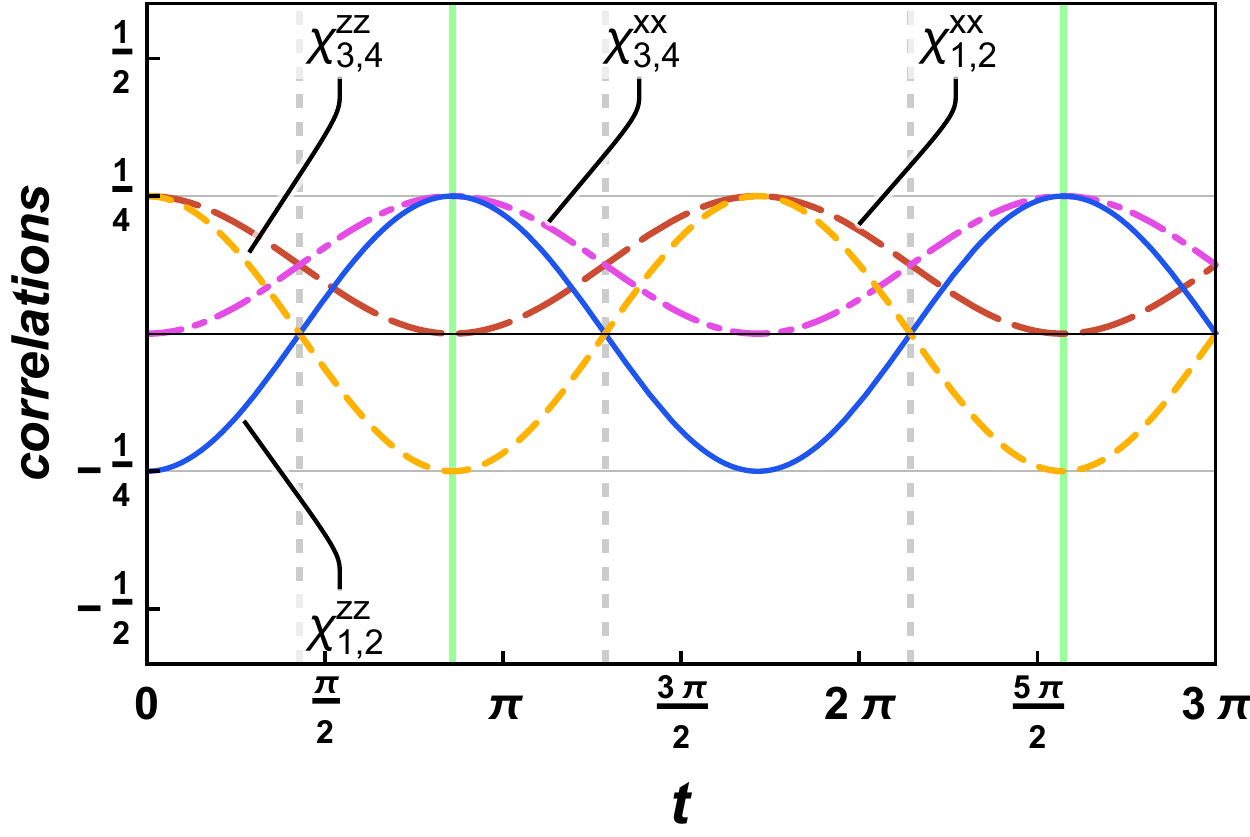}\\
		\label{fig:corr}
		\small (b)
	\end{tabular} 
	\caption{ (a)~Concurrences (b)~Correlations respectively between three and two types of pairs as a function of time for $D=0.6$. The vertical dashed gray lines present time instants of $W$ states ($t_w$) when all concurrences meet each other. The vertical solid green lines show the time instants for entanglement transition to the last pair ($t_{tr}$). Since, the concurrence can be equal to unity just for the first and last pair, the correlations between other pairs are not shown here. Also, because $xx$ correlations behaves as $yy$ ones, the first one only has been plotted.}
	\label{fig:fig2Dandcorr}
\end{figure*}
The plots~\ref{fig:Fig13D}~(a) and~(b) show the entanglement oscillation for two pairs of qubits. It is seen that at
specific time instants, which decrease in a monotonic manner with increasing $D$, $C_{1,2}$ vanishes and $C_{3,4}$ equates with unity. At the same time instants, $C_{1,3}$ gets zero (Fig.~\ref{fig:Fig13D}~(c)), as expected from concurrence relations. Hence, one can certainly infer that the entanglement completely transfers from the first initially entangled rung to the last initially disentangled one. The time of transition are
\begin{eqnarray}
t_{tr}=\frac{(2n+1)2\pi}{(\mu+\nu)},\quad\quad (n=0,1,2,\dots).
\label{t-tr}
\end{eqnarray}

\begin{figure}
	\centering
	\includegraphics[width=1.\linewidth]{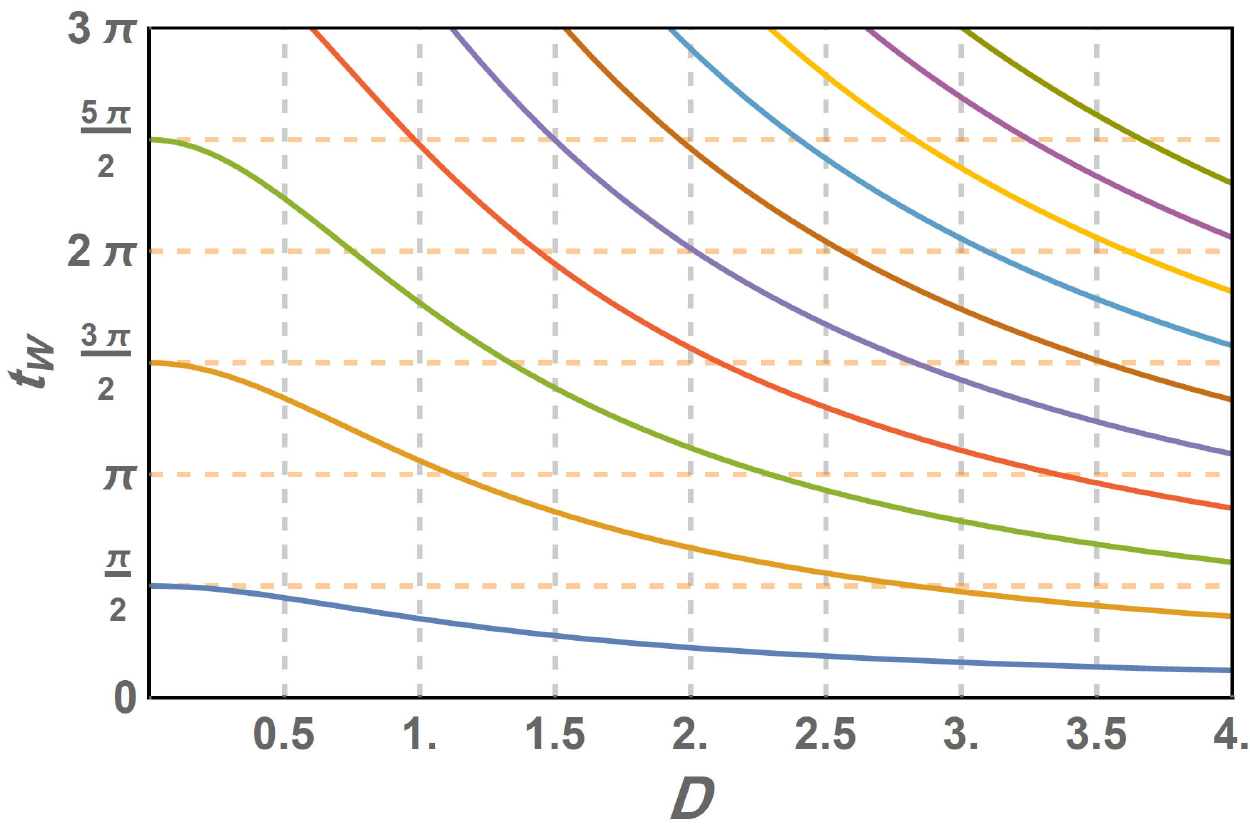}
	\caption{From the bottom ($n=0$) to the top ($n=9$), $t_w$ is shown as a function of $D$. The frequency of $W$ state generation enhances as the value of $D$ increases.}
	\label{fig:wstatestime}
\end{figure}

It turns out, when $C_{3,4}$ is equal to one, the concurrence between the other components are zero, as can be seen in Fig.~\ref{fig:fig2Dandcorr}~(a), which is a cross-sectional cut of Fig.~\ref{fig:Fig13D} for a specific value of $D$. In other words, as the entanglement turns off at one end of the spin ladder, the other end is found in the entangled Bell state.  
In addition, it is observed that at $t=\frac{(2n+1)\pi}{(\mu+\nu)}$ $(n=0,1,2,...)$, all concurrences are equal. These time instants, which henceforth are denoted by $t_w$, indicate the emerge of quantum $W$ states (see Fig.~\ref{fig:wstatestime}). 
Since the concurrences of all qubit pairs in the states of $W$ type are equal to $2/N$, then for $N=4$ (as the number of particles),  one can expect $C_{1,2}=C_{3,4}=C_{1,3}=1/2$ at $t=t_w$, as shown in Fig.~\ref{fig:fig2Dandcorr}~(a).

\section{Two-point correlations}
\label{sec:twopointcorre}
The dependence of entanglement on two-point quantum correlations~\cite{syljuasen2003} makes us explore a signature of such correlations to detect  entanglement transition and $W$ states generation.  Two-point quantum correlations ($\chi^{\alpha\alpha}_{p,q}$) between $\alpha$-th component of spin pairs ($\hat{\vec{S}}_p,\hat{\vec{S}}_q$) can be defined as
\begin{eqnarray}
\chi^{\alpha\alpha}_{p,q}(t)=\bra{\psi(t)}\hat{S}^{\alpha}_p \hat{S}^{\alpha}_q\ket{\psi(t)},
\end{eqnarray}
that is a substantial quantity. With the quantum state being given by
Eq.~(\ref{kett}) the correlation functions
have been calculated and written in table~\ref{cortable}.
\\
\begin{table}
\centering
\caption{\label{cortable}
The two-point correlations in different directions}	
\begin{tabular}{ccc}
		\hline
		Pairs Positions & $\chi^{xx}(t)=\chi^{yy}(t)$ & $\chi^{zz}(t)$ \\
        \hline		
		\rowcolor{Gray1}
		First Rung~(1,2) & $\frac{1}{4}\cos^2(\frac{\mu+\nu}{4}t)$& $\frac{-1}{4}\cos(\frac{\mu+\nu}{2}t)$\\
		Legs  & $\frac{1}{8}\sin(\frac{\mu+\nu}{2}t)$& 0\\
		\rowcolor{Gray2}
		Last Rung~(3,4) & $\frac{1}{4}\sin^2(\frac{\mu+\nu}{4}t)$& $\frac{1}{4}\cos(\frac{\mu+\nu}{2}t)$\\
		\hline
	\end{tabular}
\end{table}
Since, we seek to find the role of correlations in entanglement transition and $W$ state generation, the first and last pairs whose concurrence can take the value of one are considered in the following. For this reason, $xx$ and $zz$ correlations between $(1,2)$ and $(3,4)$ are plotted in Fig.\ref{fig:fig2Dandcorr}~(b). The comparison between Fig.\ref{fig:fig2Dandcorr}~(a) and (b), shows that when $C_{p,q}=1$, $zz$ correlation is equal to $-1/4$. On the other hand, if $C_{p,q}=0$, then such a correlation equates with $1/4$. This fact reflects the nature of the particular initial state that has been taken. Therefore, the correlation in $z$ direction can easily show the transition of entanglement. In addition, such a correlation between all pairs quenches when a $W$ state emerges as it depicted in Fig.~\ref{fig:fig2Dandcorr}~(b). 

Focusing on $xx$ ($yy$) correlation also allows us to reveal their role in dynamical behavior of entanglement. At $t=t_w$, all pairs agree on the same value of $xx$ ($yy$) correlation that is $1/8$, as shown in Fig.~\ref{fig:fig2Dandcorr}~(b) and presented in table~\ref{cortable}. In other phrase, because of quenching $zz$ correlation at $t=t_w$, just the correlations in $x$ and $y$ directions play the main role in $W$ state production. In addition, it can be said that the zero value for such correlations is in coincidence with the complete disentanglement between the considered pair. On the other hand, the maximum value of correlations in $x$ and $y$ direction is according to complete entanglement shared between that pairs.

The other types of two point correlations such as $\chi^{\alpha\beta}_{p,q}(t)$ becomes zero for all pairs, as long as $\alpha\neq\beta$. Given that, the DM interaction brings $xx$ and $yy$ correlation into existence as it is exhibited in table~\ref{cortable}, however induces a term such as $\hat{S}^{x}\hat{S}^{y}$ in the system's Hamiltonian.

In addition, as can be seen in table.~\ref{cortable}, the amplitudes of $xx$ and $yy$ correlations between leg pairs are half of that of pairs on the rungs.
The lower values for this type of amplitude, as well as the lack of $zz$ correlation between such pairs, prevent the maximum of the entanglement between them  exceeding from the value of $1/2$, which takes place at $t=t_w$.

Another important quantity is $\sum_{p=1}^{4}\hat{S}_{p}^{z}=\hat{S}_{tot}^{z}$, which does not commute with the system's Hamiltonian, nevertheless, its expectation value is temporally constant ($\bra{\psi(t)}\hat{S}_{tot}^{z}\ket{\psi(t)}=-1$). This is a generic character of $W$ states and, based on the uncertainty principle, one can reach $\bra{\psi(t)}\hat{S}_{tot}^{x}\ket{\psi(t)}=\bra{\psi(t)}\hat{S}_{tot}^{y}\ket{\psi(t)}=0$.
\section{Conclusion}
This work deals with the entanglement dynamics through a nanomagnetic triangular ladder with the Heisenberg interaction on the rungs in addition to the DM interaction over the legs, which has been treated by means of an analytical technique. In the system's initial state, the first rung is completely entangled via a Bell states and the other elements are in a separated state.  The unitary time evolution of the system's quantum state leads to the transit of entanglement from the first leg to the last one. The dependence of such transition time instants on the value of DM interaction is exactly obtained. It is found that the increasing DM interaction boosts the speed of entanglement transmission. In examining the dynamical behavior between different pairs, we arrived at the conclusion that the entanglement frequency for the legs is greater than that of the rungs. 
During the time evolution, the system passes several times through special quantum states in which one-particle states take the equal probability to be appeared. Such states whose generation is the most spectacular feature of the proposal model are so called $W$ states. 

Since the dynamical behavior of the entanglement over various pairs reflects the effect of two-point quantum correlations between them, we investigated such functions in detail. Analytical results reveal that $zz$ correlations play no role in the $W$ state generation. On the other hand, $xx$ and $yy$ correlations play the main character in the above-mentioned situations. In addition, the maximum value of such correlations ($xx$ and $yy$ ones) and the minimum value of the correlation in $z$ direction have the main role as an entanglement transition detector. 

On the other side, the interaction between a quantum system and its environment is unavoidable. For example, a quantum ladder (as the present model) might be affected by the host crystal's phonons carrying the thermal energy of the crystal at $T\neq 0 $. It is expected that the system-environment interaction will give rise to decoherence effects that can destroy the quantum correlations and can result in other effects such as entanglement sudden death~\cite{yu2009sudden} and sudden birth as reported in Ref.~\cite{mot2019}.  
\\
\section*{Acknowledgement}
M. M. is really grateful to Seyed Mohesen Rasa for drawing some figures.

\end{document}